\documentclass[english,aps,prl,twocolumn,amsmath,amssymb,showpacs,notitlepage,superscriptaddress,longbibliography]{revtex4-1}
\usepackage[T1]{fontenc}
\usepackage[latin9]{inputenc}
\usepackage{color}
\usepackage{float}
\usepackage{bm}
\usepackage{amsmath}
\usepackage{graphicx}

\makeatletter
\usepackage{amssymb}
\usepackage{amsmath}
\usepackage{graphicx}
\usepackage[colorlinks=true,linkcolor=blue,anchorcolor=red,citecolor=blue,urlcolor=blue]{hyperref}
\usepackage[caption=false]{subfig}
\usepackage{lipsum}

\makeatother

\usepackage{babel}
\begin{document}
\renewcommand{\figurename}{Fig.}
\title{Emergent Edge Modes in Shifted Quasi-One-Dimensional Charge Density Waves}
\author{Song-Bo Zhang}
\email{songbo.zhang@physik.uzh.ch}
\address{Department of Physics, University of Z\"urich, Winterthurerstrasse
190, 8057, Z\"urich, Switzerland}
\author{Xiaoxiong Liu}
\address{Department of Physics, University of Z\"urich, Winterthurerstrasse 190, 8057, Z\"urich, Switzerland}
\author{Md Shafayat Hossain}
\address{Laboratory for Topological Quantum Matter and Advanced Spectroscopy (B7), Department of Physics, Princeton University, Princeton, NJ, USA}
\author{Jia-Xin Yin}
\address{Department of physics, Southern University of Science and Technology, Shenzhen, Guangdong 518055, China}
\author{M. Zahid Hasan}
\address{Laboratory for Topological Quantum Matter and Advanced Spectroscopy (B7), Department of Physics, Princeton University, Princeton, NJ, USA}
\address{Princeton Institute for the Science and Technology of Materials, Princeton University, Princeton, New Jersey 08540, USA}
\address{Materials Sciences Division, Lawrence Berkeley National Laboratory, Berkeley, California 94720, USA}
\address{Quantum Science Center, Oak Ridge, Tennessee 37831, USA}
\author{Titus Neupert}
\email{neupert@physik.uzh.ch}
\address{Department of Physics, University of Z\"urich, Winterthurerstrasse 190, 8057, Z\"urich, Switzerland}

\date{\today}

\begin{abstract}
We propose and study a two-dimensional (2D) phase of shifted charge density waves (CDW), which is constructed from an array of weakly coupled 1D CDW wires whose phases shift from one wire to the next. We show that the fully gapped bulk CDW has topological properties, characterized by a nonzero Chern number, that imply edge modes within the bulk gap. Remarkably, these edge modes exhibit spectral pseudo-flow as a function of \emph{position} along the edge, and are thus dual to the chiral edge modes of Chern insulators with their spectral flow in \emph{momentum} space. Furthermore, we show that the CDW edge modes are stable against inter-wire coupling. Our predictions can be tested experimentally in quasi-1D CDW compounds such as Ta$_2$Se$_8$I.
\end{abstract}
\maketitle

\begin{figure}[t]
\includegraphics[width=1\columnwidth]{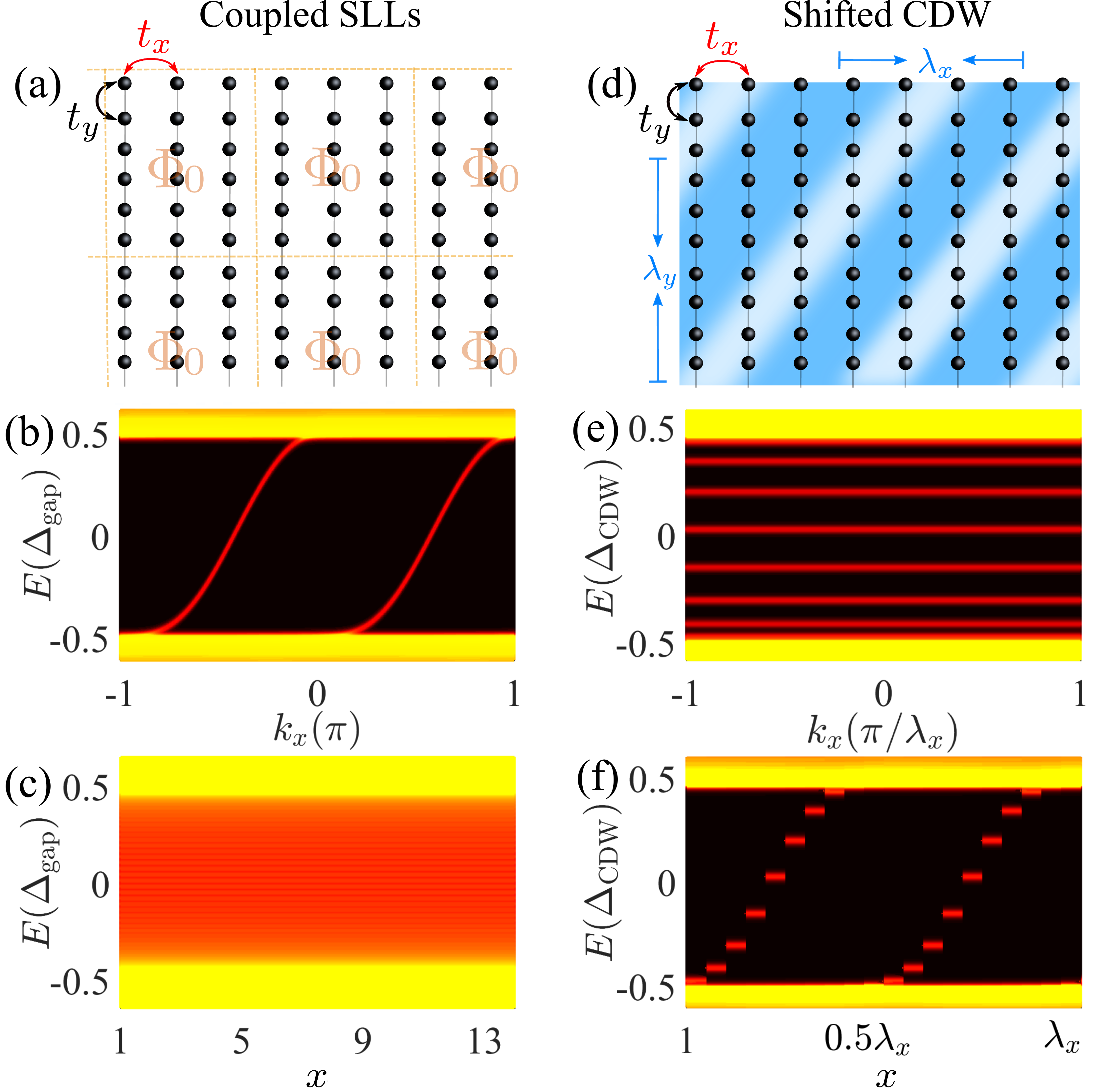}

\caption{Duality between the coupled SLLs (left) and shifted CDWs (right). (a) Schematic of the coupled SLLs. It leads to a Chern insulator. The dashed lines separate areas with unit magnetic flux $\Phi_0$. (b) Chiral edge modes (red) with energies in the bulk gap (black) of the Chern insulator, which exhibit spectral flow as a function of momentum $k_x$ along the edge. (c) Chiral edge modes in position space, which are translation invariant along the edge. (d) Schematic of the shifted CDW. The phases of the wires shift with $x$-position, as depicted by the varying background color. (e) CDW edge modes with energies inside the bulk CDW gap, which are dispersion-free in $k_x$. (f) CDW edge modes in position space, which exhibit spectral pseudo-flow as a function of $x$ along the edge.
}

\label{fig1:contrast}
\end{figure}

An insightful way to think about quantum Hall phases is in terms of an array of weakly coupled 1D sliding Luttinger liquids (SLL)~\cite{Sondhi01PRB,Kane02PRL,Mukhopadhyay01PRB,Neupert14PRB}. Each SLL consists of gapless excitations around its Fermi points $k_0\pm k_{\mathrm{F}}$, where the origin in momentum space $k_0$, is \textit{a priori} a gauge choice. When coupling identical wires that are displaced in $x$-direction and extend along $y$-direction, the difference $\delta k$ between their respective $k_0$ is an observable proportional to the flux density (i.e., perpendicular magnetic field) between them [Fig.~\ref{fig1:contrast}(a)]. A wire array built in this way is then akin to a sequence of Luttinger liquids with dispersions displaced by $k_0(x) = x\delta k$. Weak coupling between the wires opens a spectral gap in the bulk and the system enters a quantum Hall phase with chiral edge modes and quantized Hall conductivity [Figs.~\ref{fig1:contrast}(b) and ~\ref{fig1:contrast}(c)].

In this work, we contrast this construction of a topological phase from a SLL with a construction of a 2D phase of shifted charge density waves (CDW). The CDW modulation in a 1D wire can be characterized by a potential such as $\cos(Q_y y+\phi_0)$, where $Q_y$ is the CDW wavevector and $\phi_0$ is a phase that, in the case of breaking a continuous translation symmetry, is associated with the Goldstone mode of the CDW. In contrast to $k_0$ in the SLL, $\phi_0$ is not a gauge freedom of the 1D system, but determines the real-space origin of the charge density pattern. Our objective is to study the properties of an array of weakly coupled CDW wires whose phases are shifted as $\phi_0(x)=Q_x x$ [Fig.~\ref{fig1:contrast}(d)]. Remarkably, we find a duality between the edge modes of the coupled SLL and shifted CDW: On a ribbon geometry, the former has edge modes with spectral flow as a function of \emph{momentum} along the edge, while the latter has edge modes with spectral pseudo-flow~\cite{Note-pseudo-flow} as a function of \emph{position} [Figs.~\ref{fig1:contrast}(e) and \ref{fig1:contrast}(f)]. Moreover, we show that the CDW edge modes are substantially robust against inter-wire coupling.

Our study is not purely theoretically motivated, but aims to model the key aspects of the CDW compound Ta$_2$Se$_8$I. In line with the shifted CDW picture, Ta$_2$Se$_8$I consists of TaSe$_4$ chains weakly coupled by van der Waals interactions. It is known to undergo a CDW transition at $T_{\text{CDW}} \approx 260$ K~\cite{Maki83SSC,Fujishita84SSC,Requardt96JPCM,LEE85SSC,Tournier-Colletta13PRL, Gooth19Nature,Shi21Nphys,Hossain22arXiv}, developing a sizable gap that experiments determined to be between 100 and 500~meV, with a small ordering wavevector that amounts to $(Q_x,Q_z)\approx (0.054\pi/a,0.098\pi/c)$, where $c$ and $a$ are lattice constants~\cite{Shi21Nphys,HuangZL21PRB}. Recent studies highlighted a multitude of Weyl nodes that are induced through spin-orbit coupling in the low-energy electronic structure of Ta$_2$Se$_8$I above $T_{\text{CDW}}$ and their implications for possible axion physics in the CDW phase~\cite{Gooth19Nature,Shi21Nphys}. However, experimental evidence for a 3D topological (axionic) nature of the CDW state is lacking~\cite{HuangZL21PRB}. Here, we advocate a much simpler model of a shifted CDW phase for Ta$_2$Se$_8$I, for which spin-orbit coupling is unimportant. Our theory makes the experimentally testable prediction of boundary states at certain surfaces or step edges of this material.

We start by defining a minimal model on a 2D rectangular lattice in the presence of a CDW modulation, $\mathcal{H}=\mathcal{H}_0+\mathcal{H}_{\mathrm{CDW}}$ with
\begin{eqnarray}
\mathcal{H}_0 & = & \dfrac{1}{2} \sum_{\bf r} \big(t_y \Psi^\dagger_{\bf r+\hat y} \sigma_z \Psi_{\bf r}^{\vphantom{\dagger}} + t_x \Psi^\dagger_{\bf r+\hat x} \sigma_z \Psi_{\bf r}^{\vphantom{\dagger}} + h.c.\big),
\label{eq:model}
\end{eqnarray}
where $\Psi^\dagger_{\bf r}=(c_{\alpha,\bf{r}}^\dagger,c_{\beta,\bf{r}}^\dagger)$ with $c_{\sigma,\bf{r}}^\dagger$ creating an electron at orbital $\sigma\in\{\alpha,\beta\}$ and position ${\bf{r}}=(x,y)$. $\bf \hat{x}$ and $\bf \hat{y}$ are primitive vectors in $x$- and $y$-directions, respectively,  $(\sigma_x,\sigma_y,\sigma_z)$ are Pauli matrices for orbital, and $h.c.$ stands for Hermitian conjugate.
We assume that the two orbitals have hopping amplitudes with opposite signs and they do not couple via on-site or nearest-neighbor hopping terms if $x\to -x$ and $y\to -y$ mirror symmetries above $T_{\text{CDW}}$ are imposed.
This can be satisfied, e.g., when the orbitals are of $s$ and $d_{xy}$ types, respectively. Alternatively, we can rotate the Pauli matrices as $(\sigma_x,\sigma_y,\sigma_z)$$\to$$(\sigma_z,\sigma_y,\sigma_x)$ and reinterpret them in sublattice space. In this case, the two sublattices are assumed to be identical due to the mirror symmetries above $T_{\text{CDW}}$. For concreteness, we consider the orbital interpretation in the following.

The model~\eqref{eq:model} can be viewed as an array of 1D parallel wires that are displaced in $x$-direction and extend along $y$-direction. Accordingly, $t_y$ and $t_x$ are hopping strengths along and between the wires, respectively [Fig.~\ref{fig1:contrast}(d)]. Without CDW, the model is gapless with band crossings protected by the symmetries. The energy bands read $\epsilon_\pm({\bf k})=\pm (t_x\cos k_x+t_y\cos k_y)$, where ${\bf{k}}=(k_x,k_y)$ is the 2D momentum. The lattice constants are taken to be unity. We will focus on the regime $0\leqslant t_x<t_y$, such that the model approximates the relevant electronic structure of Ta$_2$Se$_8$I. The bands cross around two points (i.e., $k_y=\pm \pi/2$) along $k_y$-axis and the Fermi surfaces take ribbon shapes in $k_x$-$k_y$ plane, in agreement with those observed in Ta$_2$Se$_8$I~\cite{Tournier-Colletta13PRL,YHM21PRR,XPL21PRR}. We will consider a more realistic model for Ta$_2$Se$_8$I later.

The CDW modulation can be described as a spatially periodic local potential,
\begin{equation}
\mathcal{H}_{\mathrm{CDW}} = V \sum_{\bf r}\cos(Q_yy + Q_x x + \phi)\, \Psi^\dagger_{\bf r} \,M \, \Psi_{\bf r}^{\vphantom{\dagger}},
\label{eq:phase}
\end{equation}
where $V$ is the strength, the CDW vector $Q_y$ along each wire and the phase shift $Q_x$ in neighboring wires are related to the wavelengths $\lambda_{x(y)}$ as $Q_{x(y)}=2\pi/\lambda_{x(y)}$. We assume a limit where the wavelengths are large integers compared to the lattice constants $\lambda_{x(y)}\gg 1$~\cite{Note-rational}. $\phi$ is the global constant phase. We focus on inter-orbital CDW modulations characterized by a matrix $M\in \{\sigma_{x},\sigma_{y}\}$ which open bulk gaps at low energies, as we discuss below~\cite{Note-intra-orbital}. For illustration, we take $M=\sigma_x$, $\lambda_y=21$, $t_y=1.5$ eV and $V=0.3$ eV unless specified otherwise.

\begin{figure}[t]
\includegraphics[width=1\columnwidth]{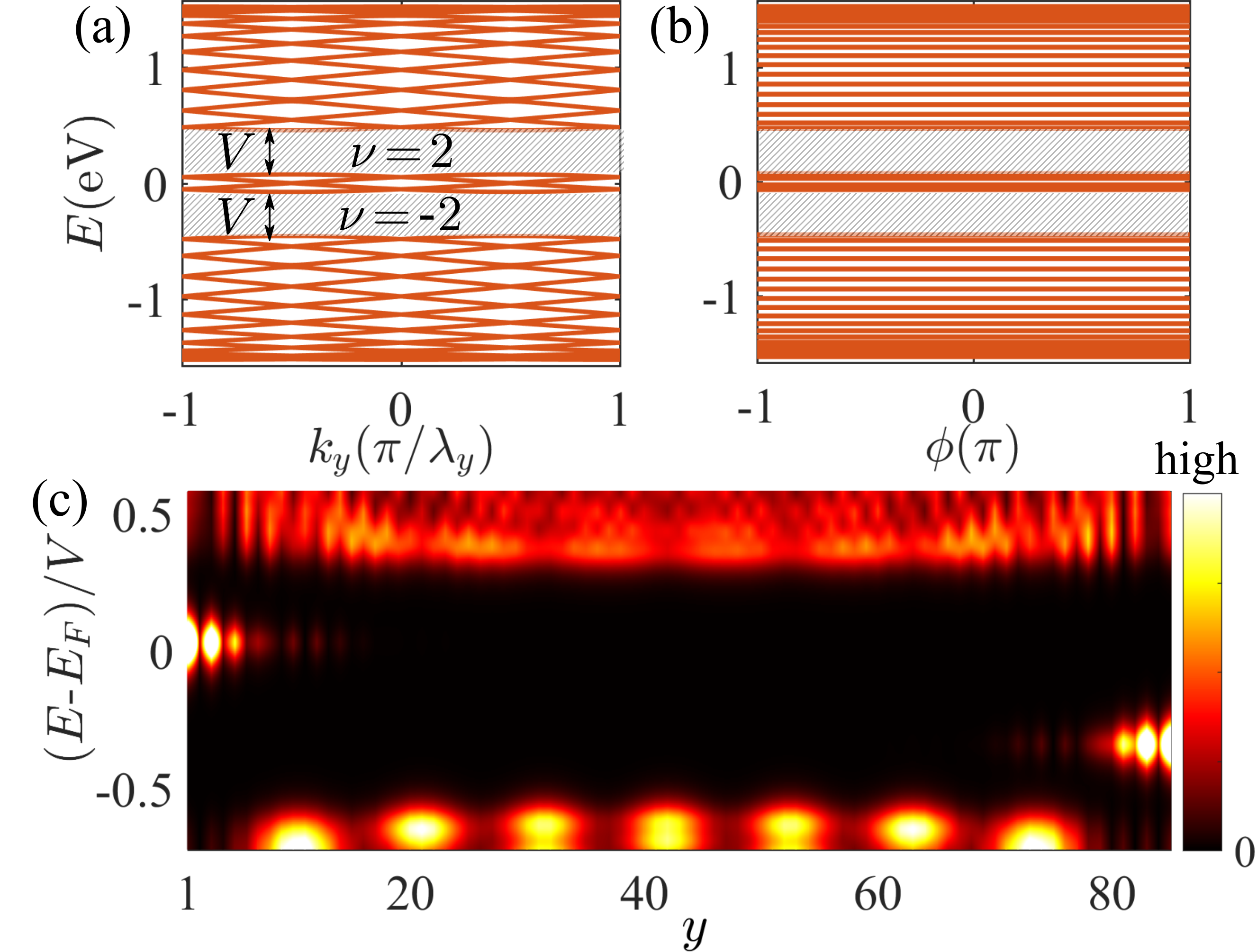}

\caption{(a) Band structure of a CDW wire with PBC. (b) Band structure as a function of $\phi$ at $k_y=\pi/(4\lambda_y)$. There are two CDW gaps of size $V$ and characterized by Chern numbers $\pm 2$. 
 (c) LDOS near the upper gap. Edge modes with energies controllable by $\phi$ appear in the gap. Parameters: $L_y=85$, $\phi_0=0.6\pi$, $E_F=t_y\sin(\pi/\lambda_y)$ and $k_BT=0.03V$ in (c).}

\label{fig2:ChernNumber}
\end{figure}

To elucidate the essential physics, we first consider the limit of decoupled wires ($t_x=0$). In this limit, all wires are identical except for their $x$-dependent CDW phases $\phi_0(x)=Q_x x+\phi$. To explore the topological properties of the system, we consider the wire at $x=0$ and impose periodic boundary conditions (PBC) in $y$-direction. Due to the super-periodic potential with large period $\lambda_{y}$, the spectrum of the wire is split into $2\lambda_{y}$ bands in the reduced Brillouin zone [Fig.~\ref{fig2:ChernNumber}(a)]. Remarkably, two bulk gaps of size $V$ emerge at $E\approx \pm t_y\sin(\pi/\lambda_y)$, respectively. The bands disperse in $k_y$, whereas are flat in $\phi$ [Figs.~\ref{fig2:ChernNumber}(a) and \ref{fig2:ChernNumber}(b)]. Note that the spectrum is periodic in both $k_y$ and $\phi$. A topological characterization of the system can be obtained in terms of Berry phase defined in the compact $k_{y}$ and $\phi$ space~\citep{LJLang12PLR}. Specifically, for each spectral gap, a Chern number can be computed as~\citep{Fukui05JPSJ}
\begin{equation}
\nu =\int_{-\pi/\lambda_{y}}^{\pi/\lambda_{y}}dk_{y}\int_{-\pi}^{\pi}d\phi\, \text{Tr}[\partial_{k_{y}}\mathcal{A}_{\phi}-\partial_{\phi}\mathcal{A}_{k_{y}}],
\label{Eq:invariant}
\end{equation}
where $\mathcal{A}_{j}=i\Phi^{\dagger}\partial_{j}\Phi$ with $j\in\{k_{y},\phi\}$ is non-Abelian Berry connection
based on the multiplet of eigenstates with energy below the gap in question $\Phi=(|\psi_{1}\rangle,...,|\psi_{m}\rangle)$.
Explicitly, we find $\nu=-2$ ($+2$) for the lower (upper) gap. These Chern numbers are independent of $\lambda_y$ for $\lambda_y\geqslant 3$.

The nonzero Chern numbers imply the appearance of midgap edge modes when open boundaries are imposed in $y$-direction, at least for a certain range of $\phi$. To illustrate this, in Fig.~\ref{fig2:ChernNumber}(c) we consider $L_y=85$ with open boundary conditions (OBC), Fermi energy $E_F=t_y\sin(\pi/\lambda_y)$, and calculate the local density of states (LDOS) as a function of position $y$ along the wire \cite{Note-EF}. Clearly, away from the boundary, the LDOS shows a bulk gap around $E_F$ that varies periodically with $y$, which is consistent with the experiments on Ta$_2$Se$_8$I~\cite{HuangZL21PRB,YHM21PRR,Hossain22arXiv}. More interestingly, inside the gap, exponentially localized edge modes appear for a wide range of $\phi$. The energies of the edge modes at opposite boundaries are generally different, and depend strongly on $\phi$, in contrast to the bulk gap that is constant in $\phi$.

\begin{figure}[t]
\includegraphics[width=1\columnwidth]{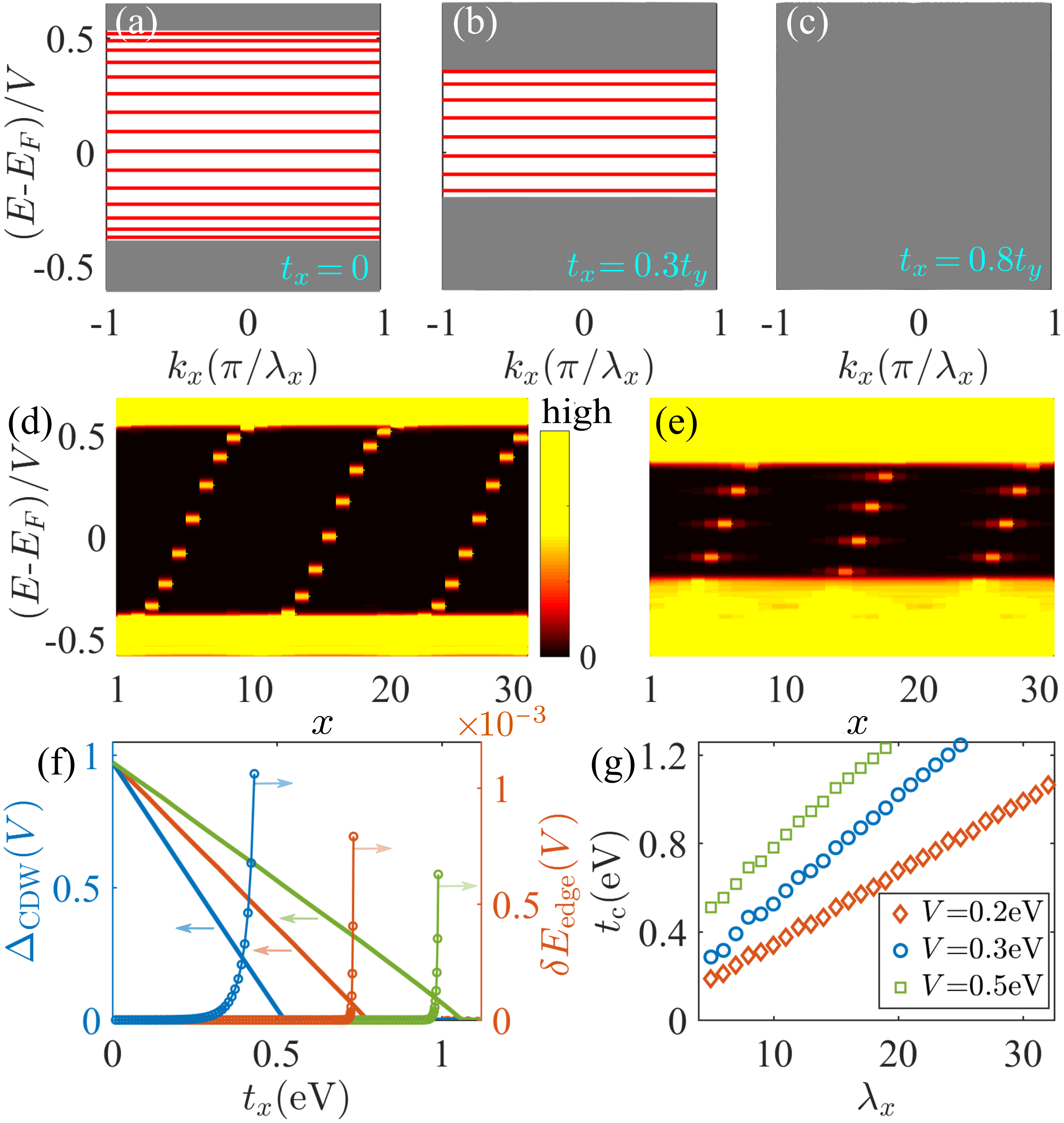}

\caption{(a)-(c) Energy spectra for $t_x=0$, $0.3t_y$, and $0.8t_y$, respectively. The bulk continuum and edge discrete spectra are indicated by gray and red color, respectively. PBC (OBC) are imposed in $x(y)$-direction. (d,e) LDOS for $t_x=0$ and $0.3t_y$, respectively. (f) $\Delta_{\mathrm{CDW}}$ (thick lines) and one edge bandwidths $\delta E_{\mathrm{edge}}$ (circle lines) as functions of $t_x$.
We consider $\lambda_x=10$ (blue), $15$ (orange) and $21$ (green) for illustration. (g) $t_c$ as a function of $\lambda_x$ for $V=0.2$ eV, $0.3$ eV and $0.5$ eV, respectively. $L_y=421$ and $\phi=0$ in all panels, $\lambda_x=21$ and $L_x=10\lambda_x$ in (d,e), and other parameters are the same as Fig.~\ref{fig2:ChernNumber}.}

\label{fig3:coupling}
\end{figure}

Crucially, in our 2D array system, the phases $\phi_0(x)$ of the wires shift in $x$-direction. The $\phi_0$ dependence of the edge modes thus implies a spectral pseudo-flow as a function of $x$ along the edge. The Chern number determines the number of pseudo-flow modes within a wavelength $\lambda_x$ along the edge. We confirm these features numerically in Fig.~\ref{fig3:coupling}(d). Moreover, in the decoupled limit, the energy spectrum of the array is flat in the reduced $k_x$ space [Fig.~\ref{fig3:coupling}(a)]. This indicates that the edge modes are immobile in $x$-direction, in stark contrast to the chiral edge modes in Chern insulators that carry current. Due to the shifted phases in the wires, at each boundary, up to $\lambda_x$ edge bands appear in $k_x$ space. Notably, the edge modes at different bands are located at different positions in each period $\lambda_x$, which again reflects the spectral pseudo-flow along the edge.

Now, we consider finite inter-wire coupling $t_x$ and show that the shifted CDW phase with the features mentioned above remain robust in the system. In Figs.~\ref{fig3:coupling}(a)-\ref{fig3:coupling}(c), we plot the energy spectra for increasing $t_x$, with PBC (OBC) in $x(y)$-direction. We find that as $t_x$ increases, the CDW gap $\Delta_{\mathrm{CDW}}$ is reduced and closed completely after a critical strength $t_{\mathrm{c}}$. Explicitly, $\Delta_{\mathrm{CDW}}$ decreases almost linearly with increasing $t_x$ [thick lines in Fig.~\ref{fig3:coupling}(f)]. For a larger $\lambda_x$, the decrease of $\Delta_{\mathrm{CDW}}$ by $t_x$ is slower and thus a larger $t_{\mathrm{c}}$ is observed [Fig.~\ref{fig3:coupling}(g)]. The critical strength $t_{\mathrm{c}}$ also increases with increasing $V$. Notably, for $\lambda_x\gtrsim 5$, $t_c$ is comparable and even larger than $V$. Due to the reduction of $\Delta_{\mathrm{CDW}}$, some edge bands are merged with the bulk continuum. Thus, the edge modes can be observed at fewer sites along the edge [Figs.~\ref{fig3:coupling}(b) and \ref{fig3:coupling}(e)]. However, the remaining edge modes with energies close to $E_F$  are only slightly extended in $x$-direction. Therefore, for large $\lambda_x$ and $V$, sizable CDW gaps with edge modes persist up to considerable inter-wire coupling in the system.

While $\Delta_{\mathrm{CDW}}$ are obviously reduced by $t_x$, the energies of edge modes remain almost dispersion-free in $k_{x}$ even for considerable $t_{x}/t_y$ as long as the edge modes persist inside the bulk gap. In Fig.~\ref{fig3:coupling}(f) we plot the width $\delta E_{\mathrm{edge}}$ of the edge band closest to $E_F$ as a function of $t_x$. We find that $\delta E_{\mathrm{edge}}$ grows as a power-law function of $t_x$. However, it is always several orders of magnitude smaller than $\Delta_{\mathrm{CDW}}$ (whose magnitude is of the same order of $V$). Overall, the flatness of edge bands against $k_x$ tends to be more pronounced for an odd and larger value of $\lambda_x$. These features can be attributed to the unique property of the spectral pseudo-flow of edge modes and that, for odd (even) $\lambda_x$, edge modes at the same energy level are separated by a distance of $\lambda_x$ ($\lambda_x/2$). The flatness of edge bands further indicates that the edge modes are immobile, even in the presence of hopping in $x$-direction.

It is important to note that the shifted CDW phase with finite $t_x$ can be characterized by the same Chern number as in the decoupled limit, since the two are adiabatically connected without a gap closure. We have also verified that the results are qualitatively the same for other forms of inter-wire coupling.  Small deformations of the model~\eqref{eq:model}, such as a deviation of the band crossing points and a difference between the hopping strengths of the orbitals, do not alter the main results~\cite{SM}.

\begin{figure}[t]
\includegraphics[width=1\columnwidth]{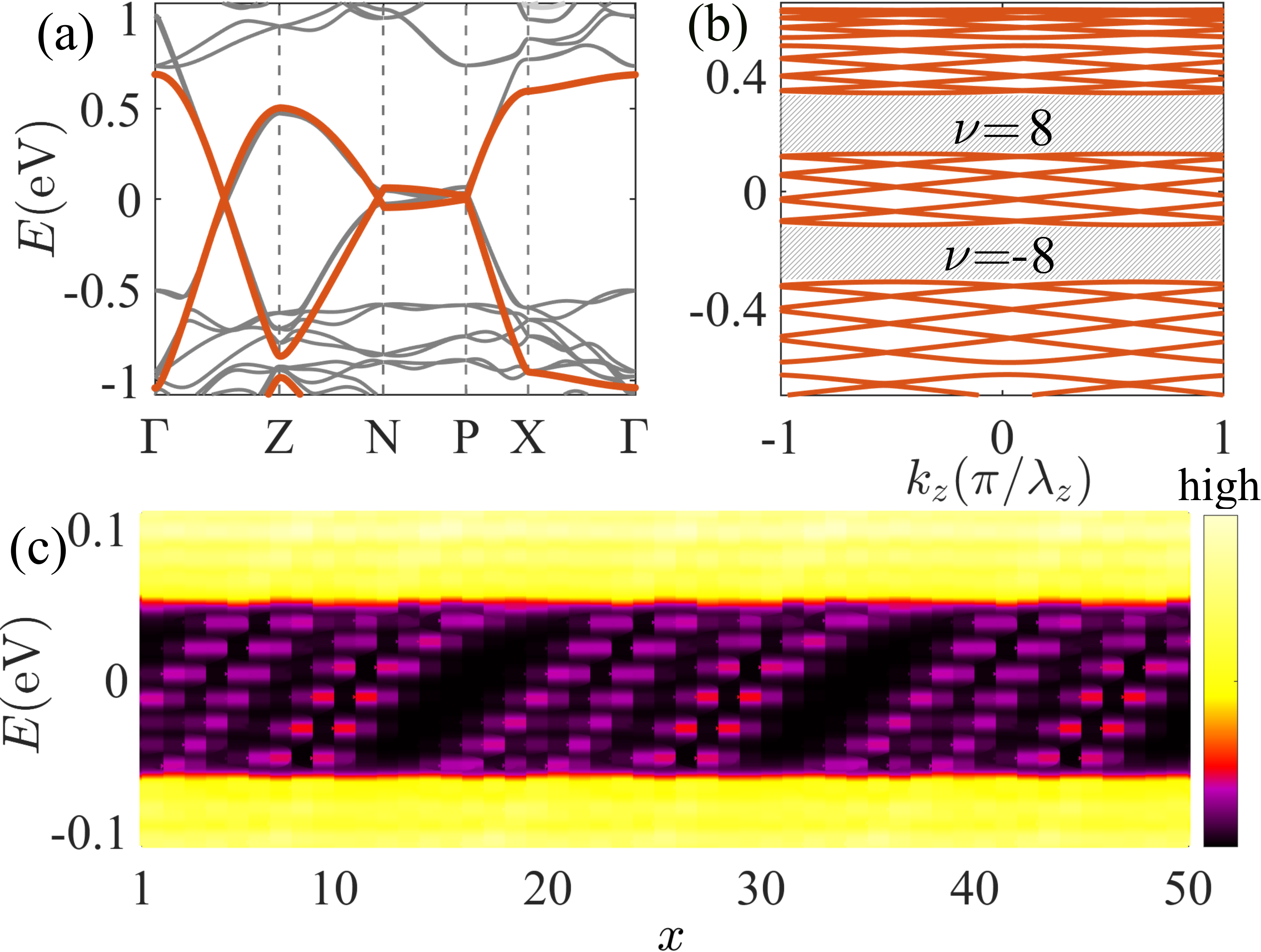}

\caption{(a) Low-energy band structure (orange) of the model~\eqref{eq:TaSeI} without CDW. Gray curves are first-principle calculations of Ta$_2$Se$_8$I. (b) Band structure with the CDW potential. Two CDW gaps, characterized by Chern numbers $\pm 8$, appear at low energies. (c) LDOS near the upper gap as a function of $\bar{x}$ (in units of $\tilde{a}$) along the edge.  Other parameters: $V=0.3$ eV, $E_F=0.25$ eV, $k_BT=0.01V$, $L_x=216\tilde{a}$ and $L_z=114c$.}

\label{fig4:TaSeI-results}
\end{figure}

 Sa far, we have shown with the minimal model~\eqref{eq:model} that the shifted CDW phase with midgap edge modes emerges in a quasi-1D system with a small CDW vector. To better connect the theory to experiment, we construct a realistic model for Ta$_2$Se$_8$I as
$\mathcal{H}_{\text{TSI}} = \mathcal{H}_{z} + \mathcal{H}_{xy}$, where the terms describing intra- and inter-chain hoppings are given, respectively, by
\begin{align}
\mathcal{H}_{z} & =\begin{pmatrix}\epsilon_{\alpha} & K_{\alpha}(k_z) & t_{1}e^{-ik_{z}c/4} & t_{1}e^{ik_{z}c/4}\\
K_{\alpha}(k_z) & \epsilon_{\alpha} & t_{1}e^{ik_{z}c/4} & t_{1}e^{-ik_{z}c/4}\\
t_{1}e^{ik_{z}c/4} & t_{1}e^{-ik_{z}c/4} & \epsilon_{\beta} & K_{\beta}(k_z)\\
t_{1}e^{-ik_{z}c/4} & t_{1}e^{ik_{z}c/4} & K_{\beta}(k_z) & \epsilon_{\beta}
\end{pmatrix},\notag \\
\mathcal{H}_{xy} & =4\cos\dfrac{ak_x}{2}\cos\dfrac{ak_y}{2}
\begin{pmatrix}0 & t_{3\alpha} & 0 & 0\\
t_{3\alpha} & 0 & 0 & 0\\
0 & 0 & 0 & t_{3\beta}\\
0 & 0 & t_{3\beta} & 0
\end{pmatrix}. \label{eq:TaSeI}
\end{align}
The model is written on the basis formed by the $d_{z^2}$-orbitals of four Ta atoms (denoted as \{$\psi_{\alpha1},\psi_{\alpha2},\psi_{\beta1},$ $\psi_{\beta2}$\}) in a unit cell. $K_{\tau}(k_z) = 2\text{Re}(t_{2\tau}e^{ik_{z}c/2})$, $\tau\in\{\alpha,\beta\}$. The parameters are given in the Supplemental Material~\cite{SM}. We can check that the model respects time-reversal, $C_{4z}$ and $C_{2x}$ symmetries in the absence of CDW. The low-energy band structure is displayed in Fig.~\ref{fig4:TaSeI-results}(a), in good agreement with first-principle calculations~\cite{YZhang20PRB}. Furthermore, considering spin-orbit coupling, the model exhibit Weyl nodes enforced by $C_{4z}$ symmetry near $E=0$.  We note that spin-orbit coupling consists of inter-chain coupling and its energy scale ($\sim$1 meV) is much smaller than that of CDW. Thus, to study the physics associated with CDW, it suffices to consider one spin species described by Eq.~\eqref{eq:TaSeI}. Similar to Eq.~\eqref{eq:phase}, we model the CDW modulation by a periodic local potential, $\mathcal{H}_{\text{CDW}} = V \sum_{{\bf r},\tau,\zeta} (-1)^\zeta \cos({\bf q}\cdot{\bf r} + \phi) \psi^\dagger_{\tau\zeta}({\bf r}) \psi_{\tau\zeta}^{\vphantom{\dagger}} ({\bf r})$, which takes opposite signs for atoms indexed by $\zeta=1$ and $2$. We choose ${\bf q} = (\pi/18\tilde{a},\pi/18\tilde{a},2\pi/19c)$, $\tilde{a}\equiv a/\sqrt{2}$ and $V=0.3$ eV, based on experimental observations~\cite{HuangZL21PRB,YHM21PRR,Hossain22arXiv}. More details about the model can be found in the Supplemental Material~\cite{SM}.

Having the realistic model, we now demonstrate that Ta$_2$Se$_8$I hosts a shifted CDW phase similar to that discussed previously. To this end, we first calculate the energy spectrum of decoupled wires (for $\mathcal{H}_{xy}=0$) under PBC in the presence of the CDW potential. As shown in Fig.~\ref{fig4:TaSeI-results}(b), two CDW gaps of size $\sim$0.2 eV appear at low energies. Using Eq.~\eqref{Eq:invariant}, we find that the gaps are characterized by Chen numbers $\nu =\pm 8$. In Fig.~\ref{fig4:TaSeI-results}(c) we take into account $\mathcal{H}_{xy}$  and consider the system on a ribbon geometry in (110) plane, with PBC (OBC) in [110] ([001]) direction. The CDW gap is reduced to be $\sim$0.12 eV. Most strikingly, inside the gap, we clearly observe edge modes (purple) with eight spectral pseudo-flows in each period ($\sim$36$\tilde{a}$).

Recently, large surface gaps with clear CDW modulations on the (110) surface of Ta$_2$Se$_8$I have been observed~\cite{HuangZL21PRB,YHM21PRR,Hossain22arXiv}. The CDW patterns have large wavelengths ($\sim$17--25 nm) both along and perpendicular to the chains. The CDW gaps ($\sim$0.1--$0.5$ eV) are smaller than the energy scale of intra-chain hopping ($\sim$1 eV), but stronger than van der Waals interaction ($\sim$0.05 eV)~\cite{Tournier-Colletta13PRL,Shi21Nphys,HuangZL21PRB,Hossain22arXiv}. These observations are in good agreement with the regime for our shifted CDW phase.
Thus, along the boundaries or step edges that are perpendicular to the chains, we predict the existence of edge modes with spectral pseudo-flow. Such crystal terminations could be prepared with focused ion beam manipulation~\cite{Moll10Nmat,Moll18ARCMP}.
We also expect our theory to be implementable in other quasi-1D CDW materials~\cite{Zawilski98SSC,ZZWang03PRB,Bolloch05PRL,Zybtsev10quantized,CBrun09PRB,LMLiu21PRL} such as TaTe$_4$ where desired CDW patterns on specific surfaces have been reported~\cite{ZhangX20PRB,SunHG20NJP}. 

Finally, we note that the edge modes can be observed even when the edge is not exactly perpendicular to the chains (but not parallel with the CDW vector).
The edge modes may be pushed into the bulk by particular discontinuous potentials at the edge. However, we expect them to be stable as long as the edge potential is smooth (i.e., the change over a lattice constant is much smaller than $\Delta_{\text{CDW}}$). Our theory can be generalized to the case with multiple CDW vectors, which we detail in the Supplemental Material~\cite{SM}.

In summary, we have proposed a 2D topological phase of shifted CDW with midgap edge modes. These edge modes exhibit spectral pseudo-flow as a function of position along the edge, thus constituting a duality compared to the chiral edge modes of Chern insulators. We have shown that this phase stays stable even under substantial inter-wire coupling. We have constructed a realistic effective model and applied the theory to Ta$_2$Se$_8$I.

\begin{acknowledgments}
We thank Claudia Felser and Glenn Wagner for fruitful discussions.
This work was supported by the European Research Council (ERC) under the European Union's Horizon 2020 research and innovation programm (ERC-StG-Neupert-757867-PARATOP) and from NCCR MARVEL funded by the SNSF. Materials characterization and the study of topological quantum properties were supported by the U.S. Department of Energy, Office of Science, National Quantum Information Science Research Centers, Quantum Science Center and Princeton University.
\end{acknowledgments}


\begin{thebibliography}{34}
\bibitem [{Note12()}]{Note-pseudo-flow}%
  \BibitemOpen
  \href@noop {} {\bibinfo  {journal} {{We use the wording pseudo-flow since the parameter
  controlling the flow -- a lattice site index along $y$-direction -- is
  not strictly continuous, while the momentum along the edge of a quantum Hall
  fluid is a continuous parameter}}\ }\BibitemShut {NoStop}%
\bibitem [{Note13()}]{Note-rational}%
  \BibitemOpen
  \href@noop {} {\bibinfo  {journal} {{We note that the calculations can be extended to the more
  general case with large commensurate (rational valued) $\lambda _{x/y}$ and
  the main results (i.e., topological CDW gaps, edge spectral pseudo-flow, and
  their stability against inter-wire coupling) discussed below remain qualitatively
  the same}}\ }\BibitemShut {NoStop}%
\bibitem [{Note14()}]{Note-intra-orbital}%
  \BibitemOpen
  \href@noop {} {\bibinfo  {journal} {An intra-orbital CDW with $M \in \protect \{\sigma
  _{0},\sigma _{z}\protect \}$ can also generate gaps which, however, are located at
  high energies (i.e., far away from the band center). In the sublattice interpretation of the Pauli matrices (with rotating ($\sigma_x,\sigma_y,\sigma_z)\rightarrow(\sigma_z,\sigma_y,\sigma_x$)), we can use a CDW modulation with opposite potentials at the two sublattices}\ }\BibitemShut {NoStop}%
\bibitem [{EF()}]{Note-EF}%
  \BibitemOpen
  \href@noop {} {\bibinfo  {journal} {{Note that $E_F$ may deviate from the crossing points in the materials, as seen in Ref.~\cite{XPL21PRR}}}\ }\BibitemShut {NoStop}%
\end{thebibliography}

%

\appendix


\setcounter{figure}{0}
\setcounter{page}{1}
\renewcommand\thefigure{S\arabic{figure}}
\renewcommand{\thepage}{S\arabic{page}}
\global\long\def\thesection{\Roman{section}}
\global\long\def\thesubsection{\Alph{subsection}}

\onecolumngrid
~\ \
~\ \
{\color{white}wwwwwwwwwwwwwwwwwwwwwwwwwwwwwwwwwwwwwwwwwwwwwwwwwwwwwwwwwwwwwwwwwwwwwwwwwwwwwwwwwwwwwwwwwwwwwwwwwwwwwwwwwwwwww
wwwwwwwwwwwwwwwwwwwwwwwwwwwwwwwwwwwwwwwwwwwwwwwwwwwwwwwwwwwwwwwwwwwwwwwwwwwwwwwwwwwwwwwwwwwwwwwwwwwwwwwwwwwwwwwwwwwwwwwwwwwww}

{\centering{\textbf{\large{Supplemental material for \\ "Emergent Edge Modes in Shifted Quasi-One-Dimensional Charge Density Waves"}}}}

~\ \
~\ \

In this Supplemental Material, we analyze the inverse participation ratio of the edge modes (Sec.~\ref{sec:IPR}), the results for other forms of inter-wire coupling (Sec.~\ref{sec:other-coupling}), under small deformations of the normal electronic structure (Sec.~\ref{sec:deformations}), the realistic effective model for Ta$_2$Se$_8$I (Sec.~\ref{realistic-model}), and the case with multiple CDW vectors (Sec.~\ref{multileCDW}).
 ~\ \
~\ \
{\color{white}wwwwwwwwwwwwwwwwwwwwwwwwwwwwwwwwwwwwwwwwwwwwwwwwwwwwwwwwwwwwwwwwwwwwwwwwwwwwwwwwwwwwwwwwwwwwwwwwwwwwwwwwwwwwww
wwwwwwwwwwwwwwwwwwwwwwwwwwwwwwwwwwwwwwwwwwwwwwwwwwwwwwwwwwwwwwwwwwwwwwwwwwwwwwwwwwwwwwwwwwwwwwwwwwwwwwwwwwwwwwwwwwwwwwwwwwwww}

\twocolumngrid

\section{Inverse participation ratio of edge modes \label{sec:IPR}}

\begin{figure}[H]
\centering
\includegraphics[width=0.68\columnwidth]{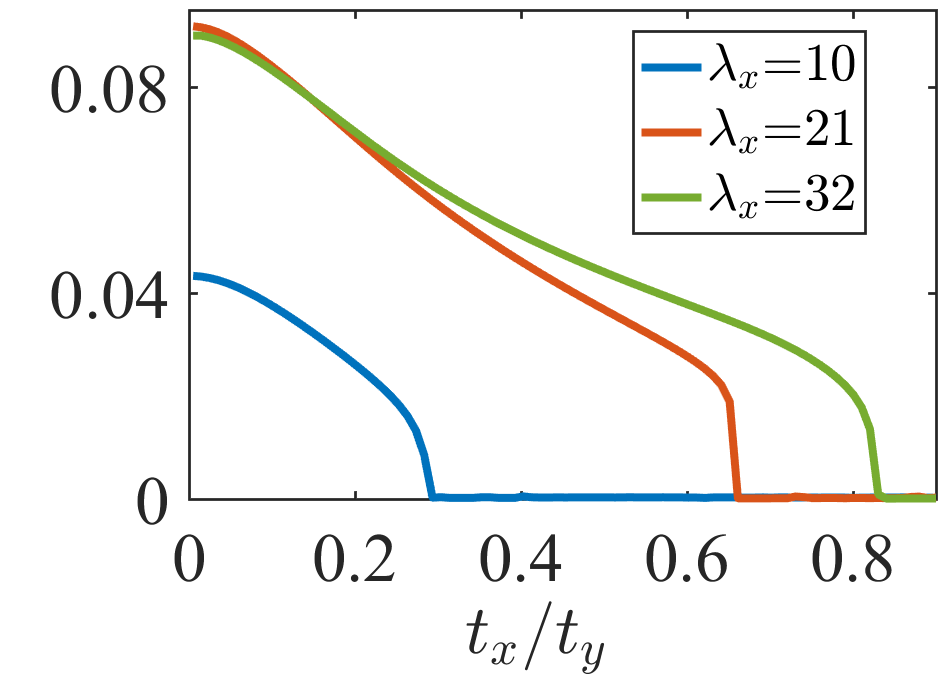}

\caption{IPR of one of the edge modes as a function of inter-wire coupling strength $t_x$ for $\lambda_x=10$, $21$ and $32$, respectively. The edge mode shows considerable IPR, as long as it persists inside the bulk gap. Other parameters are the same as Fig.~2 in the main text.}

\label{figSM0}
\end{figure}

To better show the localization properties of the edge modes in the presence of inter-wire coupling, in Fig.~\ref{figSM0}, we calculate the inverse participation ratio (IPR) of one specific edge mode with energy close to $E_F$. We find that the inter-wire coupling slightly spreads the edge mode to two neighboring sites along the edge direction, thus reducing the IPR. However, when the edge mode persists in the bulk gap, it is clear to see that the edge mode has considerable IPR. This result indicates its strong localization at the edge. For large $\lambda_x$, the IPR only vanishes abruptly when the edge mode is merging into the bulk. Similar features appear for other edge modes and they are independent of the global phase $\phi$.

~\ \
~\ \
~\ \

\section{Results for other forms of inter-wire coupling \label{sec:other-coupling}}

\begin{figure}[H]

\includegraphics[width=1\columnwidth]{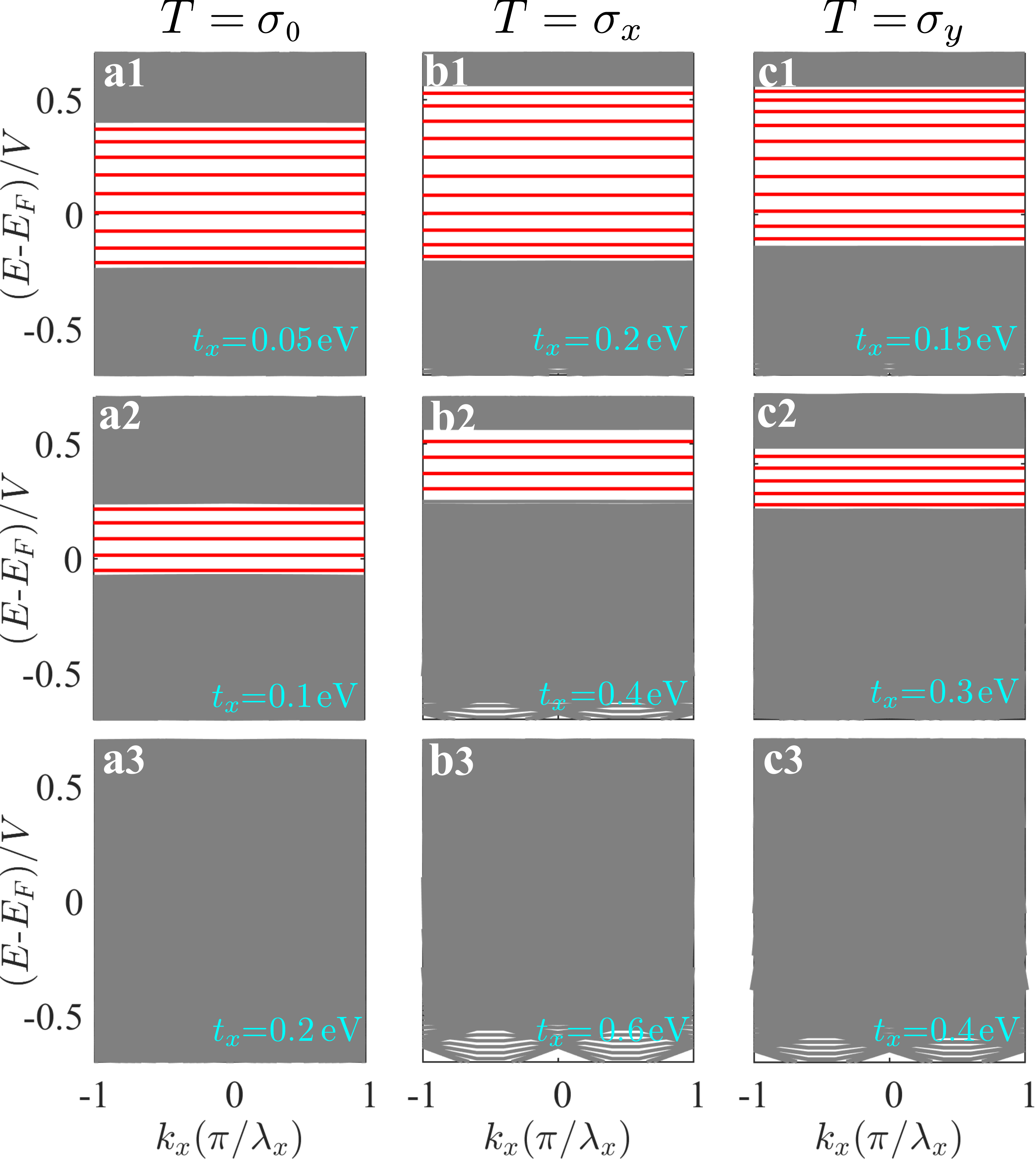}

\caption{(a1--a3) Energy spectra for $T=\sigma_0$ with $t_x=0.05$, $0.1$ and $0.2$ eV, respectively. (b1--b3) Energy spectra for $T=\sigma_x$ with $t_x=0.2$, $0.4$ and $0.6$ eV, respectively. (c1--c3) Energy spectra for $T=\sigma_y$ with $t_x=0.15$, $0.3$ and $0.4$ eV, respectively. The bulk continuum and edge discrete bands are indicated by gray and red colors, respectively. Periodic and open boundary conditions are imposed in the $x$ and $y$ directions, respectively. Other parameters are $t_y=1.5$ eV, $V=0.3$ eV, $\lambda_x=\lambda_y=21$, $L_y=421$, $E_F=t_y\sin(\pi/\lambda_y)$ and $\phi=0$.}

\label{figSM1}
\end{figure}

\begin{figure}[t]
\centering
\includegraphics[width=0.68\columnwidth]{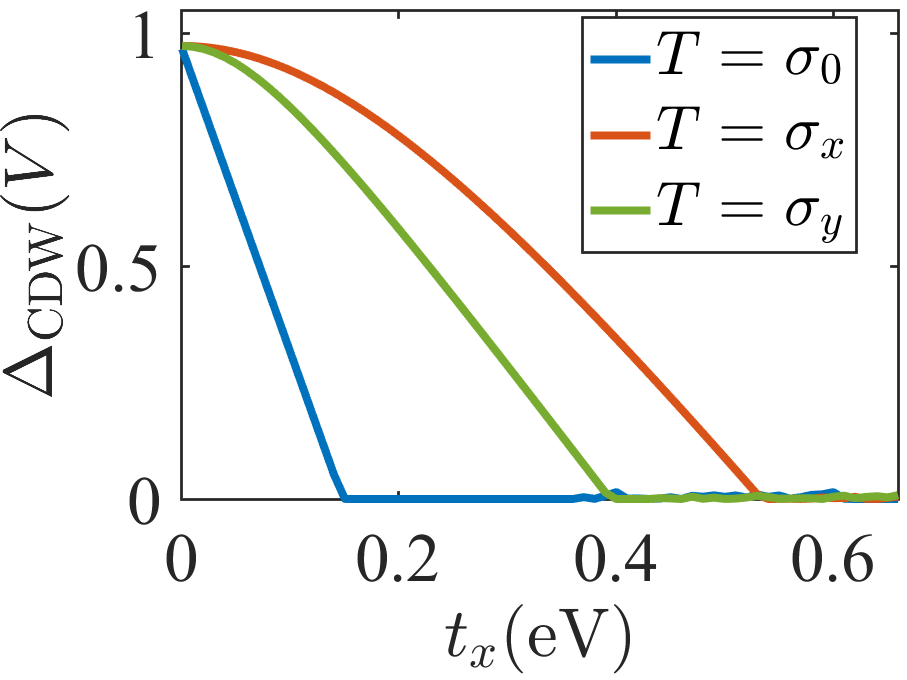}

\caption{Magnitude of CDW gap $\Delta_{\mathrm{CDW}}$ as a function of $t_x$ for $T=\sigma_0$, $\sigma_x$ and $\sigma_y$, respectively. The size $\Delta_{\mathrm{CDW}}$ decreases monotonically with increasing $t_x$ and vanishes after a large critical $t_x=t_c$. Other parameters are the same as Fig.~\ref{figSM1}}

\label{figSM2}
\end{figure}

In the main text, we discussed the case with the inter-wire coupling characterized by matrix $T=\sigma_z$. In this section, we show that the main results are more general and hold for other forms of inter-wire coupling (described by a general $2\times2$ Hermitian matrix). Note that a general inter-wire coupling may mix the two orbitals. However, for small coupling strength $t_x$, the model in the normal state remains in a metallic phase with band crossing points. A general $2\times2$ Hermitian matrix can be written as a linear combination of $\sigma_0$, $\sigma_x$, $\sigma_y$ and $\sigma_z$. To address the problem comprehensively, we present the results for the complementary cases with $T=\sigma_0$, $\sigma_x$ and $\sigma_y$, respectively (see Figs.~\ref{figSM1} and \ref{figSM2}). Similar to the main text, we choose the experimentally relevant parameters, namely, $t_y=1.5$ eV, $V=0.3$ eV and $\lambda_x=\lambda_y=21$. Clearly, in all the cases, the CDW gap is reduced by increasing inter-wire coupling strength $t_x$ and vanish after a large critical strength $t_x=t_c$. This means that the sizable CDW gap persists up to strong $t_x=t_c$ (i.e., with the magnitude in the same order as the CDW strength $V$). The edge bands remain dispersion-free in momentum $k_x$ along the edge as long as they are in the bulk gap.

\section{Results under small deformations of the normal band structure \label{sec:deformations}}

\begin{figure}[t]
\centering
\includegraphics[width=1\columnwidth]{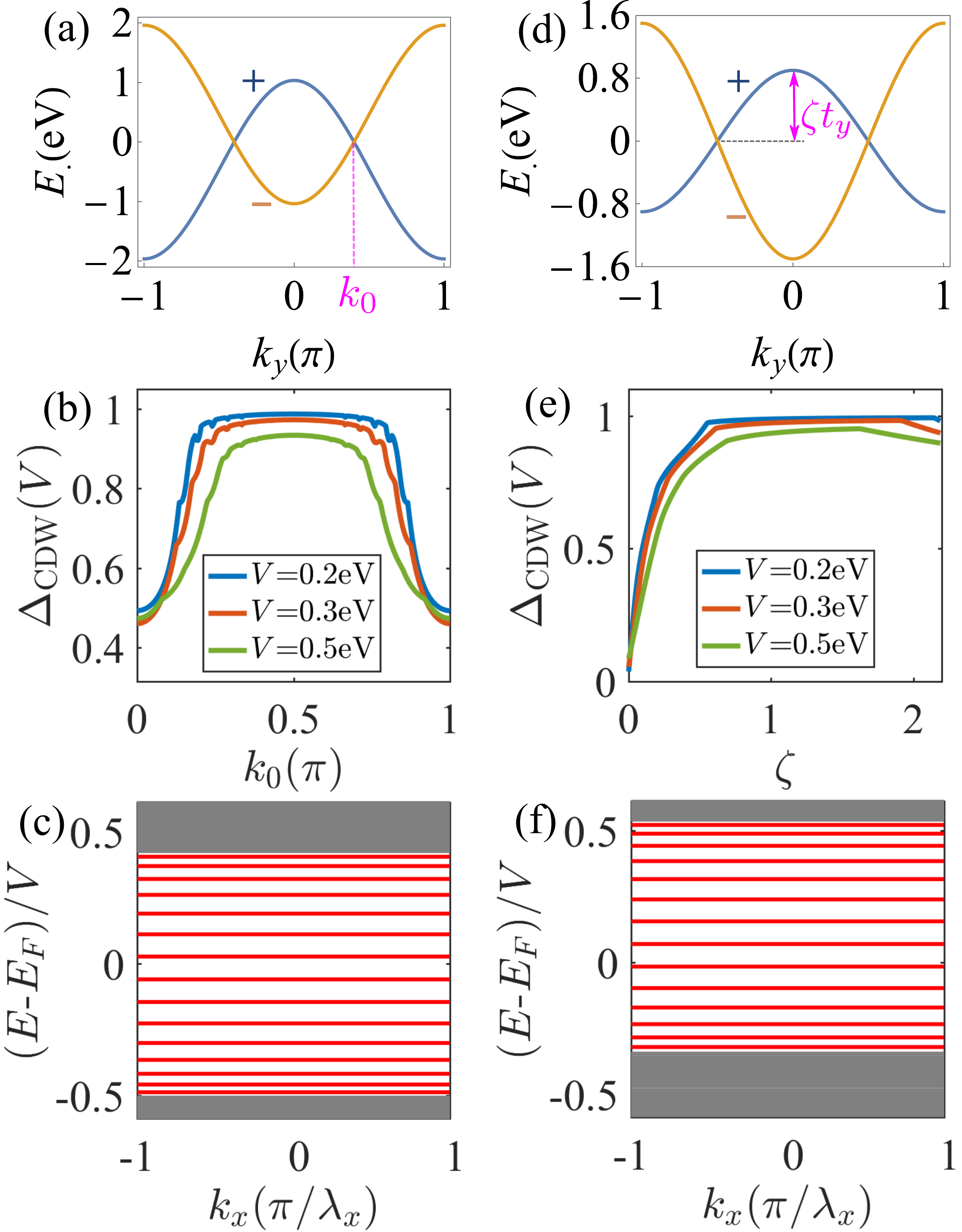}

\caption{(a) Energy dispersion in $k_y$ with the crossing points at $\pm k_0 = \pm 0.4\pi$. (d) Energy dispersion in $k_y$ for different intra-wire hopping strengths of the two orbitals. $\zeta$ denotes the ratio between the hopping strengths of the two orbitals (b) the $\Delta_{\mathrm{CDW}}$ at $E_F=t_y\sin(\pi/\lambda_y)$ as a function of the position $k_0$ of band crossing points for $V=0.2$, $0.3$ and $0.5$ eV, respectively. (e) $\Delta_{\mathrm{CDW}}$ as a function of the strength ratio $\zeta$ for $V=0.2$, $0.3$ and $0.5$ eV, respectively. (c) and (f) are the corresponding spectra of (a) and (d) with OBC in the $y$-direction and a CDW potential $V=0.3$ eV. $t_x=0$ in all plots, and other parameters are the same as Fig.~\ref{figSM1}}

\label{figSM3}
\end{figure}

In this section, we present the results for the cases where the band crossing deviated from $k_y=\pm \pi/2$ and where the two orbitals have different hopping amplitudes [see Figs.~\ref{figSM3}(a) and \ref{figSM3}(d) for the band structures], respectively. As illustrated in Figs.~\ref{figSM3}(c) and \ref{figSM3}(f), we see that the CDW gap $\Delta_{\mathrm{CDW}}$ remains roughly the same for a wide range of the deviation of the crossing point $k_0$ and difference of the hopping amplitudes $\zeta$. Inside the CDW gap, multiple edge bands appear which are localized at different sites along the edge [see Figs.~\ref{figSM3}(c) and \ref{figSM3}(f)].

\begin{figure}[t]
\centering
\includegraphics[width=0.65\columnwidth]{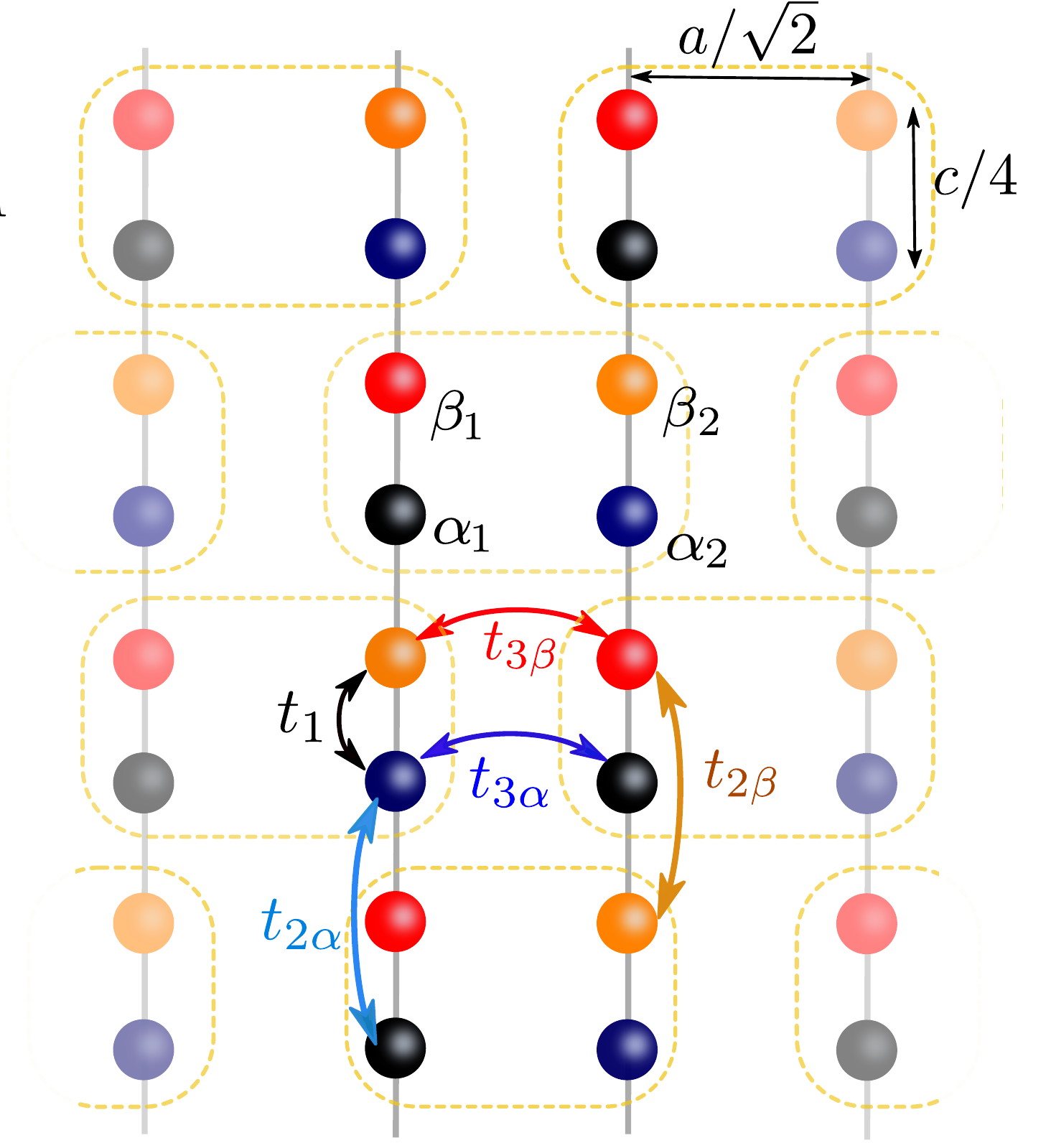}

\caption{Schematic of the lattice structure of Ta$_2$Se$_8$I in the (110) plane.}

\label{figSM-schematic}
\end{figure}

\section{Effective model for Ta$_2$Se$_8$I \label{realistic-model}}

In Ta$_2$Se$_8$I, there are four atoms in a unit cell, denoted as $\alpha_1$, $\alpha_2$, $\beta_1$ and $\beta_2$ in Fig.~\ref{figSM-schematic}. In the basis of the four atoms in a unit cell, an effective Hamiltonian can be obtained as
\begin{equation}
\mathcal{H}_{\text{TSI}} = \mathcal{H}_{z} + \mathcal{H}_{xy},
\label{eq:TaSeI-model-SM}
\end{equation}
where
\begin{align}
\mathcal{H}_{z} & =\begin{pmatrix}\epsilon_{\alpha} & K_{\alpha}(k_z) & t_{1}e^{-ik_{z}c/4} & t_{1}e^{ik_{z}c/4}\\
K_{\alpha}(k_z) & \epsilon_{\alpha} & t_{1}e^{ik_{z}c/4} & t_{1}e^{-ik_{z}c/4}\\
t_{1}e^{ik_{z}c/4} & t_{1}e^{-ik_{z}c/4} & \epsilon_{\beta} & K_{\beta}(k_z)\\
t_{1}e^{-ik_{z}c/4} & t_{1}e^{ik_{z}c/4} & K_{\beta}(k_z) & \epsilon_{\beta}
\end{pmatrix},\notag \\
\mathcal{H}_{xy} & =4\cos\dfrac{ak_x}{2}\cos\dfrac{ak_y}{2}
\begin{pmatrix}0 & t_{3\alpha} & 0 & 0\\
t_{3\alpha} & 0 & 0 & 0\\
0 & 0 & 0 & t_{3\beta}\\
0 & 0 & t_{3\beta} & 0
\end{pmatrix}, \label{eq:TaSeI}
\end{align}
and $K_{\tau}(k_z) = \text{Re}(t_{2\tau}e^{ik_{z}c/2})$, and $\tau\in\{\alpha,\beta\}$. Here, we consider only the block Hamiltonian for one spin species. The Hamiltonian for the other spin spices is the same as Eq.~\eqref{eq:TaSeI-model-SM}. The spin-orbit coupling in the full basis (including the spin degrees of freedom) can be written as
\begin{align}
\mathcal{H}_{\text {SOC}} &
=\begin{pmatrix} 0 & h_{\text {soc}} \\
 h_{\text {soc}}^\dagger & 0
\end{pmatrix}, \notag\\
h_{\text {soc}} &
=2i \begin{pmatrix} 0 & F_\alpha & 0 & 0 \\
F_\alpha & 0 & 0 & 0\\
0 & 0 & 0 & F_\beta \\
0 & 0 & F_\beta & 0
\end{pmatrix},
\end{align}
where $F_{\tau}=t_{\text{soc},\tau}^* \sin[a(k_x-k_y)/2] - t_{\text{soc},\tau} \sin[a(k_x+k_y)/2]$ with $\tau\in\{\alpha,\beta\}$. The lattice constants are $c=12.824\mathring{A}$ and $a=9.531\mathring{A}$. As illustrated in Fig.~\eqref{figSM-schematic}, $t_1$ is the hopping amplitude between nearest neighbor sites along the chains; $t_{2\alpha(\beta)}$ is the next nearest neighbor hopping associated with $\alpha(\beta)$ atoms; $t_{3\alpha(\beta)}$ is the hopping amplitude between neighboring chains; $t_{\text{soc},\alpha(\beta)}$ is the spin-orbit coupling which couples neighboring chains. These model parameters (in units of eV) are given by
\begin{align}\
\epsilon_{\alpha} & =-0.70674,\, \, \, \, \epsilon_{\beta}=-0.75538,\nonumber \\
t_{1} & =0.52768,\nonumber \\
t_{2\alpha} & =0.11988 +0.00667i,\nonumber \\
t_{2\beta} & =0.20930-0.01490i,\nonumber \\
t_{3\alpha} & =0.02188,  \
t_{3\beta} =0.02427,\nonumber \\
t_{\text{soc},\alpha} & =0.000188+0.000188i,\nonumber \\
t_{\text{soc},\beta} & =-0.00099-0.00099i.
\end{align}

In the absence of CDW, the model has time-reversal, $C_{4z}$ and $C_{2x}$ symmetries, as indicated respectively by the following relations
\begin{align}
s_y  \mathcal{H}_{tot}^*({\bf k}) s_y &= \mathcal{H}_{tot}(-{\bf k}), \notag \\
e^{i\pi s_z /4}\mathcal{H}_{tot}(k_x,k_y,k_z)e^{-i\pi s_z /4} &= \mathcal{H}_{tot}(k_y,-k_x,k_z), \notag \\
e^{i\pi s_x /2}\mathcal{H}_{tot}(k_x,k_y,k_z)e^{-i\pi s_x /2} &= \mathcal{H}_{tot}(-k_x,k_y,k_z),
\end{align}
where $s_y$ and $s_z$ are Pauli matrices acting in spin space. Moreover, it is easy to find that the model with SOC hosts Weyl points at low energies. However, we would like to note that the spin-orbit coupling is very small, compared to the CDW gap. Therefore, we ignore spin-orbit coupling and focus on one spin species in the calculations for CDW. The results for the other spin species are the same.

\section{The case with multiple CDW vectors \label{multileCDW}}

\begin{figure}[t]
\centering
\includegraphics[width=1\columnwidth]{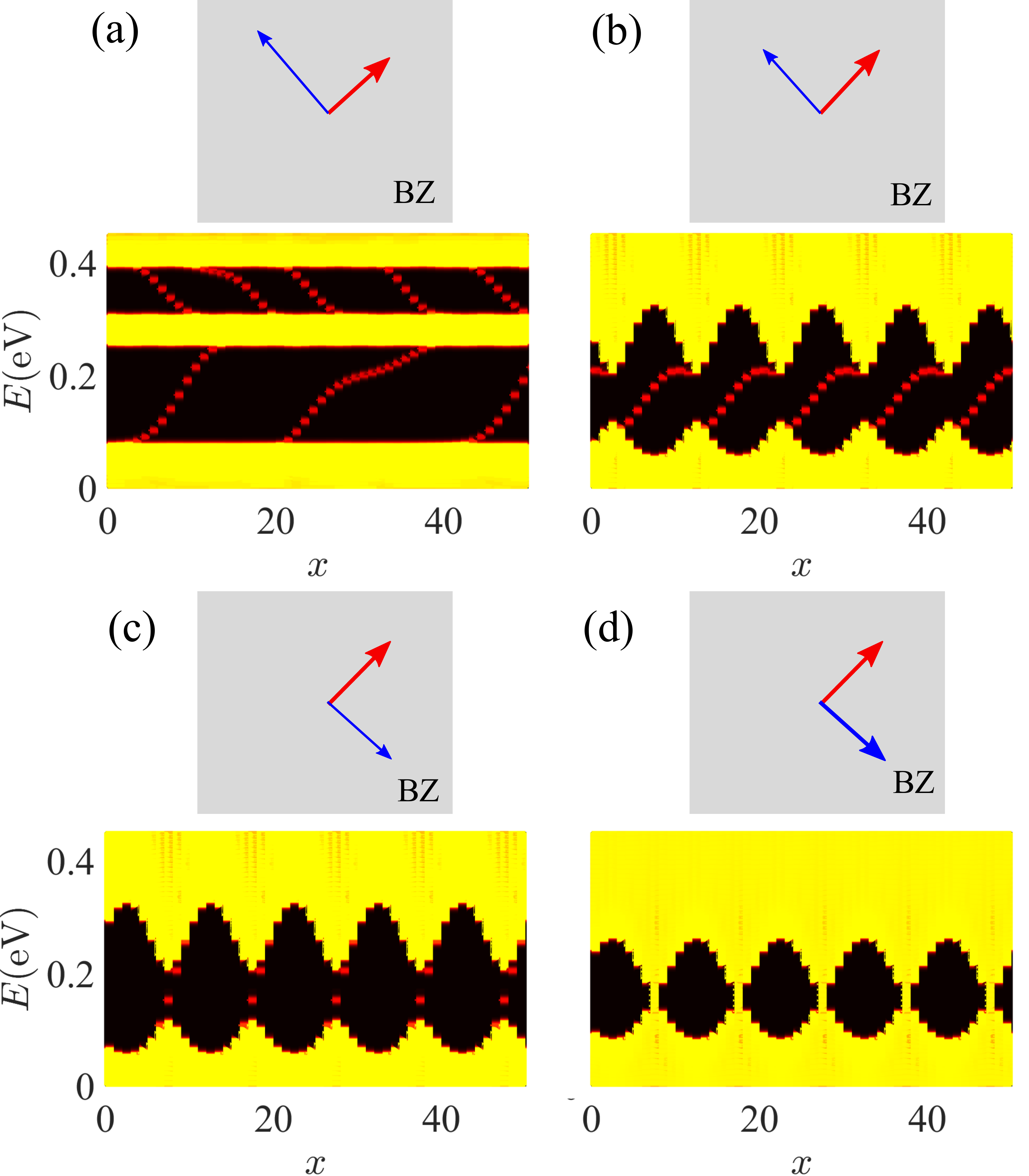}

\caption{(a) Upper: sketch of two CDW vectors (indicated by blue and red arrows) in the 2D Brillouin zone (BZ). Lower: LDOS as a function of $x$ along the edge. The two CDW vectors are ${\bf Q}_{1}=(\pi/10,\pi/20)$ and ${\bf Q}_{2}=(\pi/5,\pi/10)$. The magnitudes of the two CDW potentials are $V_1=0.2$ eV and $V_2=0.1$ eV. (b) the same as (a) but for ${\bf Q}_{1}=(\pi/10,\pi/10)$ and ${\bf Q}_{2}=(-\pi/10,\pi/10)$. (c) the same as (a) but for ${\bf Q}_{1}=(\pi/10,\pi/10)$ and ${\bf Q}_{2}=(\pi/10,-\pi/10)$. (d) the same as (c) but for $V_1=V_2=0.1$ eV.  All LDOS plotted using the minimal model.  Other parameters are $t_y=1.5$ eV, $t_x=0$, and $L_y=300$.}

\label{figSM-multipleCDW}
\end{figure}

We now discuss the case where multiple CDW vectors are present in the system.
When the CDW vectors have different wavelengths $\lambda_y$ along the wires, they each open bulk CDW gaps with nonzero Chern numbers at different energies. For the minimal model [cf. Eqs~(1,2) in the manuscript], the Chern numbers for all the gaps are given by $\nu=\pm 2$. If the wavelength perpendicular to the wires is much larger than the lattice constant, i.e., $\lambda_x\gg1$, then we can find edge modes with spectral pseudo-flows in the corresponding CDW gaps. This is shown numerically in Fig.~\ref{figSM-multipleCDW}(a).

If the 2D system has two CDW vectors with the same $|\lambda_y|$, then they open the CDW gaps at the same energies and the CDW gaps would oscillate strongly in $x$ direction [see Fig.~\ref{figSM-multipleCDW}(b-c)]. This can be understood as follows. Let us write the two CDW potentials as $U_1= V_1 \sin({\bf Q}_1\cdot{\bf r} +\phi_1)$ and $U_2=V_2 \sin({\bf Q}_1\cdot{\bf r} +\phi_2)$, where ${\bf{Q}}_i \equiv (Q_{ix},Q_{iy})= (2\pi/\lambda_{ix},2\pi/\lambda_{iy})$ and $\lambda_{1y}=\lambda_{2y}=\lambda_y$. Adding the two potentials together, we can define a net effective CDW potential as
\begin{eqnarray}\label{dd}
  U_{\text{net}} &=&  U_1+U_2 = V_{\text{net}} (x) \sin\Big[\dfrac{2\pi y}{\lambda_y} + \phi_{\text{net}}(x)\Big],
\end{eqnarray}
where
\begin{eqnarray}
   V_{\text{net}} &=& \sqrt{V_1^2+V_2^2+2V_1V_2\cos[(Q_{2x} - Q_{1x})x + \phi_2 - \phi_1)]} \nonumber \\
   \phi_{\text{net}}(x) &=& \arctan \Big[\dfrac{V_1\cos(Q_{1x} x+\phi_1)+V_2\cos(Q_{2x}x+\phi_2)}{V_1\sin(Q_{1x} x+\phi_1)+V_2\sin(Q_{2x}x+\phi_2)}\Big].\nonumber \\
\end{eqnarray}
This net CDW potential is a single periodic function in $y$-direction. Thus, at most one pairs of CDW gaps appear in the system.
Consider the weak inter-wire coupling limit ($t_x\ll {V_i,t_y}$) and that the CDW potentials $V_1$ and $V_2$ are much smaller than the intra-wire hopping $t_y$. We find that the net CDW potential $V_{\text{net}}$ oscillate between $|V_1-V_2|$ and $|V_1+V_2|$ as we move along $x$-direction. When $V_1\neq V_2$, we can still observe midgap edge modes within the bulk gaps. The form of the spectral pseudo-flows of edge modes are determined by the explicit value of $Q_{1x}$, $Q_{2x}$ and the relative phase $\phi_2-\phi_1$.

In realistic quasi-1D materials, the two CDW vectors (if they exist) should be related to each other by certain symmetries. Thus, we may expect that the CDW vectors have the same amplitudes, i.e., $V_1=V_2$ and that the two vectors fulfil the relations $(Q_{2x},Q_{2y}) = (-Q_{1x},Q_{1y})$ or $(Q_{2x},Q_{2y}) = (Q_{1x},-Q_{1y})$. In the context of quasi-1D CDW we are considering, Eq.~\eqref{dd} further indicates that there is not a full CDW gap in the 2D system.

\end{document}